\documentclass[10pt,a4paper]{article}

\usepackage{amsfonts,amssymb,amsmath}
\usepackage{amsthm,amscd}

\theoremstyle{plain}

\theoremstyle{definition}

\theoremstyle{remark}

\newcommand{\be}{\begin{eqnarray}}
\newcommand{\ee}{\end{eqnarray}}

\renewcommand{\d}{\mathrm{d}}   

\newcommand{\cf}{{\em cf.}~}

\newcommand{\Cset}{\mathbb{C}}

\newcommand{\DD}{{\mathrm D}}
\newcommand{\Det}{\operatorname{Det}}

\newcommand{\e}{{\mathrm e}}

\newcommand{\eq}{{\,  := \, }}

\newcommand{\half}{\frac{1}{2}}
\newcommand{\hol}{\operatorname{hol}}

\newcommand{\Rset}{\mathbb{R}}

\newcommand{\SO}{\operatorname{SO}}
\newcommand{\Sp}{\operatorname{Sp}}
\newcommand{\Spin}{\operatorname{Spin}}

\newcommand{\U}{\operatorname{U}}
\newcommand{\uC}{\underline{C}}
\newcommand{\uB}{\underline{B}} 
\newcommand{\ub}{\underline{b}}
\newcommand{\uA}{\underline{A}}
\newcommand{\ua}{\underline{a}}

\newcommand{\uh}{\underline{h}}

\newcommand{\Zset}{\mathbb{Z}}

\begin{document}
\begin{titlepage}
\begin{flushleft} 
\hfill Imperial/TP/041202 \\
\hfill {\tt hep-th/0412166}
\end{flushleft}

\vspace*{30mm}

\begin{center}
{\bf \Large Holonomies of Intersecting Branes}${}^{\dagger}$\\
\vspace*{8mm}

{J.~Kalkkinen} \\

\vspace*{3mm}
 
{\em The Blackett Laboratory, Imperial College} \\
{\em Prince Consort Road, London SW7 2BZ, U.K.} \\

\vspace*{6mm}

\end{center}

\begin{abstract} 

We discuss the geometry of string and M-theory gauge fields in Deligne cohomology. In particular, we show how requiring string structure (or loop space $\Spin^{\Cset}$ structure) on the five brane leads to topological conditions on the flux in the relative Deligne cohomology of the bulk - brain  pair. 

\vfill

\begin{tabbing} 
{\em PACS} \hspace{5mm} \= 11.25.-w 
{\em Email}    \>  {\tt j.kalkkinen@imperial.ac.uk} \\
\\ 
${}^{\dagger}$ Talk given in RTN-EXT Conference in Kolymbari, Greece, September 2004.
\end{tabbing}

\end{abstract}
\end{titlepage}

\section{Introduction}

Phases of wave functions carry much information that can be analysed semiclassically and that yet remains true in the full quantum theory. An example of this is charge quantisation, observed by following the phase of a wave function when it is moved around adiabatically. Similar considerations lead to flux quantisation, and a discovery of world-volume degrees of freedom that are needed to guaranteed that partition functions of branes with boundaries are just that, numbers. The arising mathematical structures involve line bundles, $n$-gerbes, bundles on branes, their loop spaces, and so on. Requiring supersymmetry in the bulk and on the brane imposes additional constraints the analysis of which is the main purpose of this contribution.  These constraints can be expressed in terms of various cohomology theories, Deligne cohomology most notably, and we shall see, in particular, in which sense the topological class of the 11-dimensional four-form flux in M-theory and the world-volume three-form on the five-brane determine a class in relative Deligne cohomology of the bulk - brane pair. 

For concreteness, let us see some of these ideas in action and 
consider a quantum mechanical particle, or brane, with world-volume $W$, embedded in the ambient space-time $X$. Move it adiabatically to an other location $W' \subset X$ along some trajectory $\Sigma$ such that $\partial \Sigma = W' - W$. The wave function then changes, typically, by a phase that depends on $\Sigma$
\be
\psi' &=& \hol(\Sigma) \psi ~. 
\ee
For an electrically charged particle in the Maxwell field $F =\d A$ one observes
\be
\hol(\Sigma) = \exp i q_{e} \int_{\Sigma} F = \e^{ i q_{e} \int_{W'} A} \cdot  \e^{ -i q_{e} \int_{W} A} ~. 
\ee
This means that the wave function must be of the form
\be
\psi = \e^{i S_{\text{inv}} + i q_{e} \int_{W} A} ~. 
\ee
In the presence of a monopole at $Q \subset X$ the Bianchi identity is violated by $\d F = q_{m} \delta_{Q}$. If we move the electric charge around a closed path $\Sigma$ from $W$ back to itself $W'=W$, we get the consistency condition 
\be
1 = \e^{ -i q_{e} \int_{W'} A} \cdot  \e^{ +i q_{e} \int_{W} A} =  \e^{ i q_{e} \int_{\Sigma} F } =  \e^{ i q_{e} \int_{D} \d F } = \e^{ i q_{e}q_{m} \int_{D} \delta_{Q} } ~, 
\ee
where $\partial D = \Sigma$ and $D$ intersects $Q$ once. 
This leads of course to the Dirac-Zwanziger quantisation condition $q_{e}q_{m} \in 2\pi\Zset $.

\section{Geometry of wave functions} 

Given a background flux $F$ such that $\d F = 0$ we can always find a local gauge potential $F|_{i} = \d A_{i}$ on a contractible chart ${\cal U}_{i} \subset X$. The standard coupling $\hol(W) \sim \exp \int_{W} A$ must be modified to take this into account. The precise formula is
\be
\psi \sim \hol_{\uA}(W) &:=& \exp(\sum_{e} \int_{e} A_{e} + \sum_{v \in \partial e} \ln \lambda_{ve}) ~,  
\ee
where we have divided the path $W$ into edges $e$ that lie each completely in some chart denoted, accordingly, ${\cal U}_{e}$. This requires, however, that the pair $\uA \eq [A^{[1]}_{i},\lambda_{ij}^{[0]}]$ satisfy 
\be
-A_{i} + A_{j} + \d \ln \lambda_{ij} &=& 0 \qquad \text{   on }   {\cal U}_{ij} \\
\lambda_{ij}\lambda_{ik}^{-1}\lambda_{jk} &=& 1 \qquad \text{   on } {\cal U}_{ijk} ~. 
\ee 
Let us introduce the shorthand notation $\DD \uA =0$ 
to denote these two cocycle conditions, and $\d \uA = F$.  
The cocycle conditions remain satisfied even if we redefine $\uA \longrightarrow \uA + \hat\DD \uh$ where
$ \hat\DD \uh = (\d \ln h_{i}, h_{i}h_{j}^{-1})$ 
by virtue of $\DD \hat\DD = 0$. The gauge equivalence class of  $\uA \sim \uA + \hat{D}\uh$ determines a class \cite{brylinski} in the \v{C}ech-de Rham cohomology $[\uA] \in  H^{2}(\underline{\Omega}^{0}(X, \Cset^{*}) \longrightarrow \underline{\Omega}^{1}(X, \Cset) )$. Keeping track of $\ln \lambda_{ij}$ rather than just the phase $\lambda_{ij}$, we get a representative of smooth Deligne cohomology $ H^{2}_{\cal D} := H^{2}((2\pi i)^{2} \Zset \longrightarrow \underline{\Omega}^{0}(X, \Cset) \longrightarrow \underline{\Omega}^{1}(X, \Cset) ) $. These classes determine an isomorphism class of line bundles with connection. 

We could similarly consider $ \uB = [B_{i}, \eta_{ij}, g_{ijk}]  \in 
\underline{\Omega}^{0}(X, \Cset^*)
\stackrel{\d\log}{\longrightarrow} \underline{\Omega}^{1}(X,
\Rset) \stackrel{\d}{\longrightarrow} \underline{\Omega}^{2}(X,
\Rset) $ and impose $D  \uB =0$, where  $D=\d-\delta$, modulo $\uB
\sim  \uB + \hat{D}\uA$. The cocycle conditions $D  \uB =0$ can be expanded
\be
1 &=& g_{jkl} g_{ikl}^{-1} g_{ijl} g_{ijk}^{-1} \\
0 &=&  -\eta_{ij} - \eta_{jk} + \eta_{ik} + g_{ijk}^{-1} \d g_{ijk} \\
0 &=& B_{i} - B_{j} + \d \eta_{ij}
\ee
This defines a class in $H^{3}_{\cal D}$, which represents an isomorphism class of one-gerbes with connection and curving. We can go arbitrarily high in degree by defining 
\be
H^{n}_{\cal D} &:=& H^{n}\Big( (2\pi i)^{n} \Zset \longrightarrow \underline{\Omega}^{0}(X, \Cset) \longrightarrow \cdots \underline{\Omega}^{n-1}(X, \Cset) \Big) ~. 
\ee

The holonomy of $\uB$ around a surface $\Sigma$ can be written down in terms of a triangulation of the surface $t = \{s,e,v\}$
\be
\hol_{\uB}(\Sigma) &:=&  \exp(\sum_{s} B_{s} - \sum_{e \subset \partial s} \int_{e} \eta_{es} + \sum_{v \in \partial e} \ln g_{ves}) ~,
\ee
where $s$ indexes sides, $e$ edges and $v$ vertices of this triangulation. Note that also a choice of assigning a chart ${\cal U}_{i}$ to each simplex is involved. 
The holonomy has two notable properties: 
\begin{itemize}
\item[-] Consider $\Sigma = \partial V$ for some three-manifold $V$. Since $H \in H^{3}(X,2\pi i\Zset)$ 
\be
\hol_{\uB}(\partial V) &=& \exp \int_{V} H 
\ee
does not depend on the choice of $V$. For $H=0$, in particular, the mapping $\hol_{\uB} : H_{2}(X) \longrightarrow \U(1)$ is a representation of $H_{2}(X)$ in $\U(1)$. 
\item[-] The holonomy depends on the representative of the Deligne class as follows
\be
\hol_{\uB + \hat\DD \uA}(\Sigma) &=& \hol_{\uB}(\Sigma) \hol_{\uA}(\partial\Sigma)~. \label{ltrans}
\ee
\end{itemize}
These results generalise naturally to Deligne classes of arbitrary degree.

\section{Loop space geometry}

The world-volume theory on the five-brane involves spinors, locally, so one might expect that the five-brane should be chosen in such a way that it has spin structure. Spin$^{\Cset}$ structure on $M$ is not enough, because there are no vector multiplets at our disposal in an $N=(0,2)$ theory. What we do have at our disposal, however, is 
\begin{itemize}
\item[-] A Spin$^{\Cset}$ structure on the loop space ${\cal L}M$, often referred to as string structure. 
\item[-] A $\Spin_{G}$ structure on $M$.
\end{itemize}
Let us look at these two structures one by one. 

{\bf String structure}. It turns out \cite{gawedzki} that we can view the last term in (\ref{ltrans}) as a gauge transformation on a line bundle evaluated at $\partial\Sigma \in {\cal L}X$ in loop space and that $\hol_{\uB}(\Sigma)$ is really a section of the line bundle. This line bundle with a connection is determined  by the one-gerbe with curving and connection $\uB$, and changes of equivalence class by $\hat{D}\uA$ correspond to standard gauge transformations. Loosely speaking, the connection at $\gamma \in {\cal L}M$ is ${\cal A}_{i} \sim \oint_{\gamma}B_{i} + \cdots$ and the transition functions ${\cal G}_{ij} \sim \oint_{\gamma}\eta_{ij} + \cdots$. 
This structure amounts to a Spin$^{\Cset}$ structure on the loop space ${\cal L}M$. 

Requiring Spin$^{\Cset}$ structure on the loop space boils down to requiring that the structure group of the loop space spin bundle have a central extension. This structure is often called {\em string structure} \cite{Killingback:1986rd}. Suppose, in particular, that $P \longrightarrow M$ is a principal $\SO(V)$ bundle and $Q \longrightarrow M$ its lift to a principal $\Spin(V)$ bundle, $\dim V \geq 4$. If $M$ is simply connected then the loop space ${\cal L}M$ is connected. Then also the principal bundle ${\cal L}P \longrightarrow {\cal L}M$ can be lifted to a ${\cal L}\Spin(V)$ bundle ${\cal L}Q \longrightarrow {\cal L}M$. 
We say that ${\cal L}Q \longrightarrow {\cal L}M$ has {string structure} if the structure group ${\cal L}G \eq {\cal L}\Spin(V)$ can be extended by a circle 
\be 
1 \longrightarrow S^{1} \longrightarrow \widehat{{\cal L}G} \longrightarrow {\cal L}G \longrightarrow 1 ~. 
\ee 
This happens if $\lambda(P) = 0$; the converse is true if $\pi_{2}(M) = 0$ \cite{dml}.   

{\bf Spin${}_{G}$ structure}. The scalars (and the spinors in a certain sense) transform under the structure group of the normal bundle $\SO(5)$. In the tensor theory this group is lifted to its double cover $G=\Sp_{2} = \Spin(5)$. We may then ask for fibre bundles 
\be
\Sp_{2} \longrightarrow \Spin_{G}(1,5) \longrightarrow \SO(1,5)  ~, \label{ext}
\ee
and require $\Spin_{G}$ structure, \cf \cite{Avis:1979de}. Note that this structure arises automatically when pulling back spinors from the bulk onto the five-brane. One can indeed verify in the orientable case $w_{1}(M)=0$ that spin structure in the bulk $w_{2}(X)$ forces the obstructions to a global $\Sp_{2}$ bundle and a global $\Spin(1,5)$ bundle to cancel. The necessary  condition for this extension to exist \cite{jk} is $W_{5}(TM) =0$, which implies $w_{4} \equiv \lambda' \mod 2$ for some class $\lambda' \in H^{4}(M; \Zset)$.

\section{Branes ending on branes and cohomology}

Let us fix a submanifold $M \hookrightarrow X$ and the inclusion $\iota : M \longrightarrow X$. Relative cohomology of the pair $(X,M)$ classifies closed cocycles $\d \Omega =0$ on $X$ whose pullback by $\iota$ determines a trivial class $\iota^{*} \Omega = \d \omega$, for some $\omega$, in the cohomology of $M$. These classes can be conveniently given in terms of the pair $(\Omega, \omega)$ that satisfies
$\d (\Omega, \omega) \eq (\d \Omega, \iota^{*}\Omega - \d \omega) \equiv 0 $. 
The coboundary operator is nilpotent $\d^{2}$. Two pairs define the same cohomology class if they differ by a locally exact term
$(\Omega', \omega')  - (\Omega, \omega) = \d (\Lambda, \lambda)$.   
The same idea should work in relative \v{C}ech-de Rham cohomology as well. We define 
\be
\DD (\uB, \ua) & \eq & (\DD \uB, \iota^{*}\uB - \DD{\ua} ) \\
\d (\uB, \ua) & \eq &  (\d \uB, \iota^{*} B - \d \ua) ~.
\ee
One should point out that the notation implies that $\iota^{*}\uB - \DD{\ua}$ has no two-form piece, and that $B$ stands for the local representative $B_{i}$ when a chart ${\cal U}_{i}$ is specified. Again $\hat\DD = \DD + \d$, $\DD\hat\DD \equiv 0$, and $\d^{2}\equiv0$. 

The bosonic open string partition function is a number, not a  section of a nontrivial line bundle. To formulate it properly, we have to fix boundary conditions $\partial \Sigma \subset Q$, and take degrees of freedom on the D(irichlet) brane into account. These degrees of freedom include the Chan-Paton gauge field $\ua = [a_{i}^{[1]}, \lambda_{ij}^{[0]}]$. As $\DD \ua \neq 0$, these one-forms and circle-valued functions do not define a line bundle with connection globally, but do so only locally. The obstruction to this is precisely the $\uB$-field as $\DD \ua = \iota^{*}\uB$. 
The structure here is relative Deligne cocyle $(\uB,\ua)$, and the bosonic open string partition function is of the form
\be
\psi_{\text{tot}}(\Sigma,\partial\Sigma) &=& \e^{-S_{\text{inv}}}\hol_{\uB}(\Sigma) \hol_{\ua}(\partial\Sigma)^{-1}~. 
\ee
This is now well-defined under the equivalence  $(\uB, \ua) \sim  (\uB, \ua) + \DD (\uA, \underline{h})$.

Smooth Deligne cohomology group, $H^{p}_{\cal D}(X)$, and the group of differential characters \cite{CS} of degree $p$, $\hat{H}^{p}(X, (2\pi i)^{p} \Zset)$, are (canonically) isomorphic \cite{brylinski}. Let us see now how this works in the present context: The wave function $\psi_{\text{tot}}(\Sigma,\partial\Sigma)$ is a differential character as long as the relative field strength $(H,f)$ belongs to $H^{3}(X,Q; 2\pi i \Zset)$. To verify this, suppose $(\Sigma,\partial\Sigma) = \partial (V,d)$: then 
\be
\hol_{\uB}(d - \partial V) \hol_{\ua}(\partial d)^{-1} &=& \exp -\int_{(V,d)} (H,f) 
\ee
which is independent of the choice of $(V,d)$ precisely when 
the flux quantisation in relative cohomology applies. Precisely in the same way one would expect the M2-brane wave function to be
\be
\psi_{\text{tot}}(W,\partial W) &\sim& \e^{-S_{\text{inv}}}\hol_{\uC}(W) \hol_{\ub}(\partial W)^{-1}~, 
\ee
where $G = \d \uC$ is the four-form field strength in 11-dimensional supergravity and $\ub$ is the world-volume two-form on a M5-brane with self-dual field strength. Due to the interactions that are present already in the bosonic theory, and the fact that the world-volume string is self-dual, the situation is somewhat more complicated \cite{Witten:1996hc}, and the above coupling, in the form it arises, involves factors of half \cite{Kalkkinen:2002tk}.

Superstring partition functions involve world-volume fermion determinants $\Det \d_{\Sigma}$ that can be thought of as sections of certain line bundles on the loop space as well. In particular, as shown in \cite{FW}, if we move the open string $(\Sigma, \partial\Sigma)$ over the volume $(V,d)$ such that $\partial(V,d) =0$ then 
\be
\Det \d_{\Sigma} &\longrightarrow & (-)^{\int_{d}w_{2}(Q)} \Det \d_{\Sigma} ~. 
\ee  
As the partition function should still be a number and not a section, this contribution changes the cocycle condition to $\DD(\uB,\ua) = (0,w_{2}) $, where $w_{2} \in H^{2}(Q,\Zset_{2})$ is the second Stiefel-Whitney class \cite{FW}. In local form, the scalar component of the cocycle conditions changes to $\iota^{*}g_{ijk} (\lambda_{ij}\lambda_{jk}\lambda_{ki})^{-1} = (w_{2})_{ijk} $. 
On the level of Chan-Paton flux $F|_{i} = \d a_{i}$, this boils down to the twisted quantisation condition 
\be
\left[ \frac{F}{2 \pi i}\right] -\half x& \in& H^{2}(Q, \Zset) ~,
\ee
where $x = c_{1}(L)$ for a line bundle $L$ such that $x^{2}$ is odd.  

The four-form flux $G \eq \d \uC$ in the bulk obeys a similar shifted quantisation condition \cite{Witten:1997md}
\be
\left[ \frac{G}{2 \pi i}\right] -\half \lambda& \in& H^{4}(X, \Zset) ~,
\ee
where $\lambda = p_{1}(TX)/2$ is the canonical four-class of the spin 11-manifold $X$. This arises from the membrane world-volume determinants, and can be accounted for -- at least in the orientable case \cite{Witten:1997md} -- by setting $\DD \uC = w_{4}$. (This observation was discussed independently in Ref.~\cite{Aschieri:2004yz}.) We want to see now if this obstruction to defining a global $\uC$ field can be accommodated in the relevant relative variant of Deligne cohomology as was the case with Freed-Witten anomaly:  For this, let us assume now that 
\begin{itemize}
\item[-] The 11-dimensional bulk $X$ is spin and orientable. 
\item[-]
The five-brane world-volume is either compact and orientable or it is of the form $M_{5} \times \Rset$ where $M_{5}$ is compact. 
\item[-] $M$ is simply connected and has a string structure in the sense below.
\item[-] $M$ has a Spin${}_{G}$ structure with $G=\Sp_{2}$.
\end{itemize} 
The last assumption should follow by construction from embedding $M$ into the spin manifold $X$. 

A well-defined quantum theory of non-critical strings on $M$ should require that the generalised spin bundle ${\cal L}\Spin_{G}(1,5)  \longrightarrow {\cal L}M$ can be extended centrally. As in the standard ${\cal L}\Spin(1,5)$ case, obstructions to this reside in $H^{2}( {\cal L}M, \underline{S}^{1}) \simeq H^{3}({\cal L}M,\Zset)$; what still remains to be shown is in the image of what characteristic classes in $H^{4}(M,\Zset)$ they are. As $\Spin_{G}(1,5)$ is not simple, the central extensions are by a torus rather than a circle. If we require a central extension similar to the standard spin case,  the obstruction should be the same $p_{1}(TX)/2 =0 $ \cite{jk}. 

It follows now from Wu's formulae that $\iota^{*}w_{4}(TX) = w_{4}(TM)$. As the obstruction for Spin${}_{G}$ structure should vanish $W_{5}(TM)=0$, we see that $w_{4}(TM) \equiv \lambda(TM) \mod 2$. As the quantum structure of the loop space ${\cal L}M$ requires precisely $\lambda(TM)=0$, the pullback of $w_{4}(X)$ on the five-brane $M$ is trivial $\iota^{*}w_{4} = 0$. The arising twisted cocycle condition is therefore  $\DD(\uC,\ub) = ( w_{4}(X), v_{3}(M) )$ 
for some closed $v_{3}$. Note that the above analysis does not imply $v_{3}=0$.  Determining $v_{3}$ precisely would require repeating the analysis of \cite{FW} in six dimensions for the quantum loop space structure of the self-dual string moving on the five-brane $M$.

\section{Conclusions}

The five-brane tensor multiplet that contains the world-volume two-form $\uB$ provides a natural $\Spin^{\Cset}$ structure in the loop space of the five-brane. This structure is represented by the class $[{\cal A}_{i}, {\cal G}_{ij} ] \in H^{2}_{\cal D}({\cal L}M)$. The topological obstruction, often referred to as string structure, is that $\lambda(TM) = 0$ should vanish  as a differential form. Under mild restrictions on topology, the vanishing of this obstruction guarantees that the 11-dimensional field $\uC$ determining an Abelian bulk two-gerbe (with connection and other structure) and the world-volume field $\ub$ determining an Abelian world-volume one-gerbe (with structure) have the relative topological charge
\be
\DD(\uC,\ub) &=& \Big( w_{4}(X), v_{3}(M) \Big) ~. 
\ee 
This is to be contrasted to the Freed-Witten anomaly that can in this context put in the form
\be
\DD(\uB,\ua) &=& \Big( 0, w_{2}(Q) \Big) ~. 
\ee 
It would be interesting to determine the class $v_{3}(M)$ as a six-dimensional non-critical generalisation of the Freed-Witten anomaly in ten-dimensions. Note that that very anomaly on a D6-brane can be directly related to the shifted flux of the M-theory four-form as was shown in \cite{Sparks:2003ck}. 

The structure of the M-theory three-form is one of the intriguing current geometrical issues. One way to model it is in terms of $E_{8}$ gauge fields and twisted Deligne classes \cite{Diaconescu:2003bm}. This leads indeed to an elegant description of the M-theory wave functions \cite{Freed:2004yc} in the absence of five-brane sources.

\subsubsection*{Acknowledgements}
Supported by Particle Physics and Astronomy Research Council (PPARC) Postdoctoral Fellowship.

\end{document}